\documentstyle[preprint,aps,epsf]{revtex}
\tighten
\begin{document}
\newcommand{\rb}[1]{\raisebox{-1ex}[-1ex]{#1}}
\newcommand{\sst}[1]{\scriptscriptstyle{#1}}
\newcommand{\dfrac}[2]{\mbox{$\displaystyle\frac{#1}{#2}$}}

\draft

\title{
Realistic Ghost State : 
Pauli Forbidden State from Rigorous Solution of the $\bbox{\alpha}$ Particle
}

\author{H.~Kamada$^{
\dagger }$\footnote{present address: 
Institut f\"ur Strahlen- und Kernphysik der Universit\"at  Bonn
Nussallee 14-16, D53115 Bonn, Germany    },
         S. Oryu   and  A. Nogga$^{\dagger}$
}
\address{
Department of Physics, Faculty of Science and Technology,
Science University of Tokyo, Noda,  Chiba 278, Japan} 
\address{
$^\dagger$ 
Institut f\"ur  Theoretische Physik II,
         Ruhr-Universit\"at Bochum, 44780 Bochum, Germany}


\date{\today}
\maketitle

\begin{abstract}
The antisymmetrization of the composite particles in cluster model calculations manifests 
itself in Pauli forbidden states (ghost states), if one restricts oneself on undeformed cluster
in the low energy region.
The resonating group method and the generating coordinate method 
rely on a property of the norm kernel, which introduces  
some of the  ghost states. 
The norm kernel has been usually calculated under
the assumption that the inner wavefunctions have a simple Gaussian form.
It is the first time that this  assumption is tested by the rigorous way.
In the ${}^4$He+N system, we demonstrate a ghost state,   
which is calculated from a rigorous solution of Yakubovsky equations for 
the  $\alpha$ particle.
The ghost states calculated  by rigorous and approximate methods   
turn out to have a very similar form. 
It is analytically proved that the trace of the norm kernel does not 
depend on the inner wavefunction we choose.
\end{abstract}

\narrowtext

Since 1937\cite{Wheeler}, 
when the resonating group method (RGM) was established, it has been 
 successfully  applied to many   
 light nuclei systems.
The method is essentially based on the variational principle under the 
conditions that the clusters remain in  the ground state 
in the low energy region, and the total wave function is  totally
antisymmetrized because of the Pauli exclusion principle.  
A typical example is the  two $\alpha$ model\cite{Tang,Kircher} of  ${}^8$Be.  
In 1970's the RGM had great successes\cite{Ikeda} and 
the method was extended to the generating coordinate method (GCM)\cite{GCM}, 
the orthogonality condition model (OCM)\cite{OCM}, 
the fish-bone optical model (FBOM)\cite{FBOM}, etc.   
The Pauli exclusion principle plays an important role in the relative motion
part of the wave function because it rules out part of the model space by an orthogonality condition. 
The Pauli forbidden states (ghost states) are generated by diagonalization of an integral kernel in the RGM. 
The integral kernel is known as the norm kernel (NK) ${\cal N}$  
which is defined, for example, in the two cluster model as  
\begin{eqnarray}
{\cal N}(\vec r, \vec {r'})  \equiv
 < \phi_1 \phi_2 \delta (\vec r) \vert (1 - {\cal A} ) 
\vert \phi_1 \phi_2 \delta (\vec {r'}) > 
\label{NK}
\end{eqnarray}
where $\phi_1 $, $\phi_2$ and ${\cal A}$ are two inner cluster  
wavefunctions of  the system and an antisymmetrizer for all nucleons, 
respectively. $r$ is the relative motion coordinate.  
Namely, the ghost states $u_n (\vec r)$ are eigenstates of $\cal N$ with the eigenvalue 
$\gamma_n=1$. 
\begin{eqnarray}
\int d \vec {r'} {\cal N}(\vec r, \vec {r'}) u_n (\vec {r'}) =
 \gamma_n u_n(\vec r)
\label{eigen}
\end{eqnarray}
The total wavefunction $\Psi$ of  the system 
\begin{eqnarray}
\vert \Psi > \equiv {\cal A} \vert \phi_1 \phi_2 \chi > 
\end{eqnarray}
is orthogonal to the ghost states  $ \vert \phi_1 \phi_2 u_n > $ 
\begin{eqnarray} 
< \Psi \vert \phi_1 \phi_2 u_n > = < \phi_1 \phi_2 \chi \vert {\cal A} 
\vert \phi_1 \phi_2 u_n > =   < \phi_1 \phi_2 \chi \vert (1 -1 ) 
\vert \phi_1 \phi_2 u_n >  =0,
\end{eqnarray}
where $\chi $ is the relative wavefunction between the two clusters.

Beyond ${}^8$Be the $\alpha$ cluster model has been studied 
in  ${}^{12}$C\cite{Kamimura,Fujiwara,Kamada}, 
${}^{16}$O\cite{ORYU}, etc. 
The correct treatment of the Pauli forbidden states is essential even in the 
case of bound states of clusters, where the neglect of the Pauli principle 
leads to an extreme overbinding. However, even with the condition they are over-bound, 
which is still a pending problem\cite{3alpha-overbinding,Kamada-3clusterforce,Kato} in the 
$\alpha$ cluster model. 

It is analytically proved\cite{Zaikin} that if the inner wavefunctions 
are  simple products of Gaussian functions, then the eigenvectors 
$u_n$ of (\ref{eigen}) become familiar harmonic oscillator functions.
For the sake of simplicity four spinless
cluster ($\alpha$ particle) system in $^{16}$O has been studied\cite{ORYU} 
using the Yakubovsky equations\cite{Yakubovsky}. 
Nowadays, it is possible to obtain rigorous solutions of the $\alpha$ particle wavefunction
using realistic potentials\cite{Kamada-Gloeckle}. 
Recent Progress of 4N scattering state is reviewed in\cite{Fonseca}.
Therefore, it becomes possible to compare the ghost states which one obtains from 
the rigorous Yakubovsky solution of the $\alpha$ particle.

In this letter we would like to choose the most simple case, 
the $\alpha$ + N system.

\begin{figure}[hbtp]
\begin{center}
\mbox{\epsfysize=40mm\epsffile{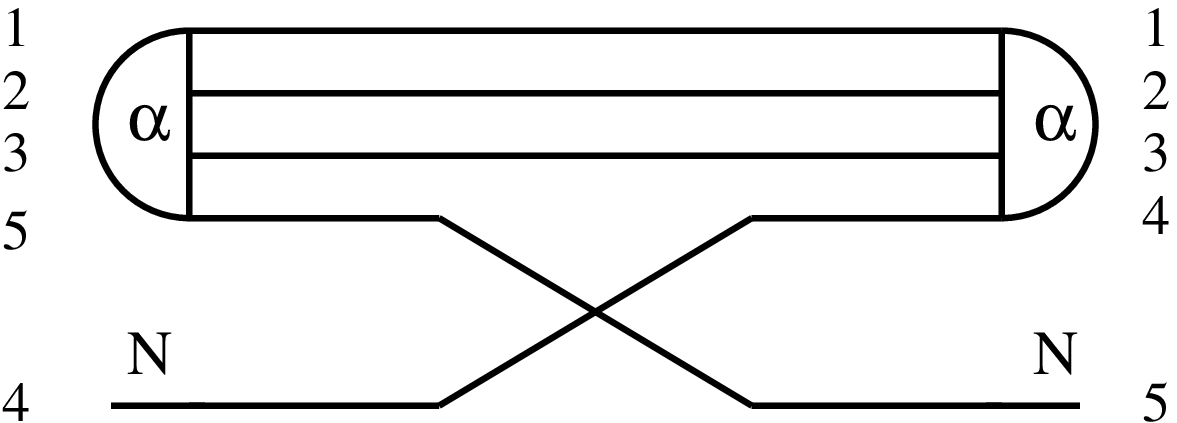}}
\caption{ The diagram of the norm kernel. 
}
\label{FIG1}
\end{center}
\end{figure}

\begin{figure}[hbtp]
\begin{center}
\mbox{\epsfysize=40mm\epsffile{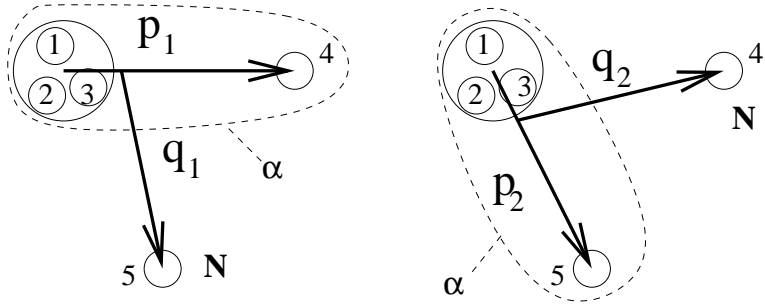}}
\caption{ Jacobi momenta }
\label{FIG2}
\end{center}
\end{figure}

The first 4 nucleons build up the ground state of the $\alpha$ particle, while the fifth particle
is the spectator.
The nucleons are 
identical particles, furthermore, the wavefunction
 $\phi_\alpha$ of 
$\alpha$ particle is  normalized to 
$< \phi_\alpha \vert \phi _\alpha > = 1 $
and NK is defined as  Eq. (\ref{NK}),
\begin{eqnarray}
{\cal N}  = 4 < \phi_\alpha \phi_n \delta \vert P_{45} \vert 
\phi_\alpha \phi_n \delta >.
\label{norm}
\end{eqnarray}
$P_{45} $
is the particle exchange operator (4 and 5).
Fig.\ref{FIG1} suggests the picture  of three cluster system (3N + N +N).
The calculation of the NK is done with Jacobi coordinates 
which we show in Fig. \ref{FIG2}.  
The relative Jacobi momenta  are prepared  and 
the relations between the Jacobi coordinates are 
\begin{eqnarray}
\vec p_1 = \vec q_2 + {1 \over 4} \vec q_1 , \,\,\,\,
\vec p_2  = \vec q_1 + {1 \over 4} \vec q_2
\end{eqnarray}
where $p_i$ and $q_i$ ($i=1,2$) are Jacobi momenta of 3N+N relative motion  
and 4N + N one, respectively.

Our way of calculating the NK is very  similar to the calculation of a leading (Born)
term of the Alt-Grassberger-Sandhas equations\cite{AGS,TEX-Gloeckle}.
For example, in the text\cite{Thomas} they treat the Born term $Z$ by the 
partial wave representation with a function $F^{\cal L} $.
In our calculation it is simply replaced as 
\begin{eqnarray}
F^{\cal L}_{N_1 N_2} (q_1,q_2) = { 4 \over 2} \int _{-1} ^1
d \cos \theta  
\left[  \int_0 ^\infty d x \int_0 ^\infty  dy  
\phi_{\alpha : N_1} (x, y, \vert \vec  p_1 \vert ) 
\phi_{\alpha : N_2} (x, y, \vert  \vec p_2 \vert)  \right]
P_{\cal L}(\cos \theta) 
\end{eqnarray}
where the angle $\theta$ is between the vectors $\vec q_1 $ and
$\vec q_2 $,  $N_1$ and $N_2$  the state channels of the partial wave, and 
$P_{\cal L} $  the Legendre function. $x$ and $y$ are rests of Jacobi momenta
which describe the motion of particles (1,2 and 3) inside of the $\alpha$ 
particle. The numerator $4$ derives from (\ref{norm}). 

As an example, the $\alpha$ wavefunction\cite{Kamada-Gloeckle} 
of the Argonne potential (AV14)\cite{AV14} is applied, and 
we take the case of total spin J = 1/2$^+$. 
For the sake of simplicity we assume the spin $j$ of 3N  is almost 1/2$^+$ 
(in fact, 94.9\% for the case of AV14 potential ), 
therefore, the angular momentum between  clusters $\alpha$ and neutron
is S-wave. This leads to $\cal L$=0.

Under this choice of the partial waves the recoupling coefficient  $A^{\cal L}_{N_1,N_2}$ 
\cite{Thomas} is 1/4 and one gets
\begin{eqnarray}
{\cal N}_0 ( q_1, q_2) = { 1 \over 2} \int _{-1} ^1 d \cos \theta \left[ \int dx \int dy \phi _{ \alpha : [1/2^+]} 
(x,y,p_1) \phi_{\alpha : [1/2^+] } (x,y,p_2) \right] \equiv { 1 \over 2} 
\int _{-1} ^1 d \cos \theta \tilde { \cal N } ( p_1 , p_2 ) 
\end{eqnarray}
where the subscript ``0'' of the norm kernel means the angular momentum of $\vec q$ and the kernel
 $\tilde {\cal N}$ will be used later.

\begin{figure}[hbtp]
\begin{center}
\input{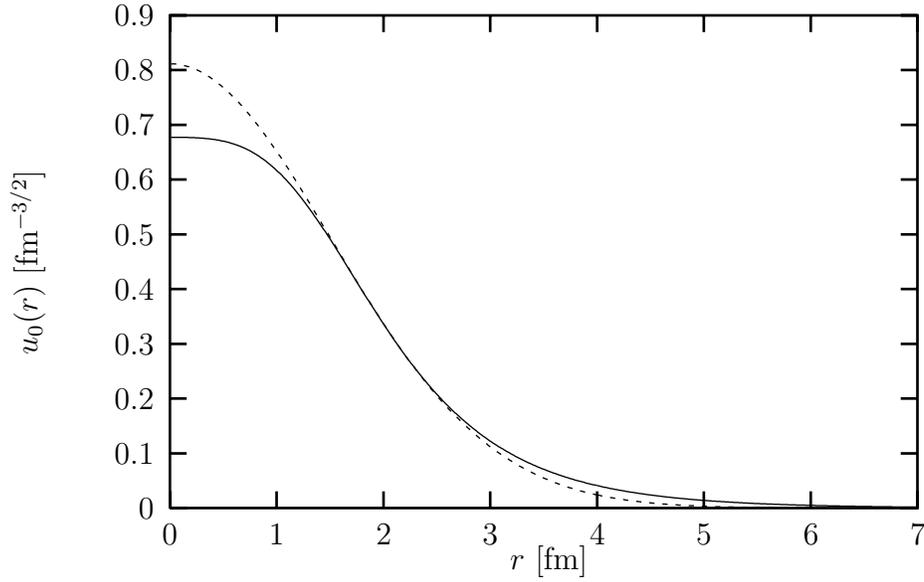}
\caption{ The ghost state in coordinate space.
The solid (dashed) line is $u_0^Y(r)$ ($u_0^G(r)$).
}
\label{FIG3}
\end{center}
\end{figure}

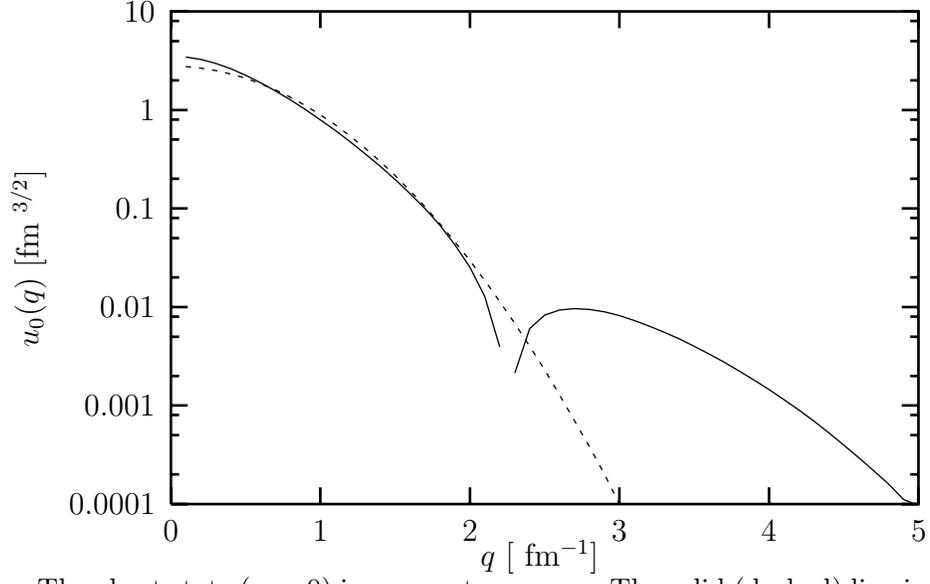
\begin{figure}[hbtp]
\begin{center}
\setlength{\unitlength}{0.1bp}
\special{!
/gnudict 40 dict def
gnudict begin
/Color false def
/Solid false def
/gnulinewidth 5.000 def
/vshift -33 def
/dl {10 mul} def
/hpt 31.5 def
/vpt 31.5 def
/M {moveto} bind def
/L {lineto} bind def
/R {rmoveto} bind def
/V {rlineto} bind def
/vpt2 vpt 2 mul def
/hpt2 hpt 2 mul def
/Lshow { currentpoint stroke M
  0 vshift R show } def
/Rshow { currentpoint stroke M
  dup stringwidth pop neg vshift R show } def
/Cshow { currentpoint stroke M
  dup stringwidth pop -2 div vshift R show } def
/DL { Color {setrgbcolor Solid {pop []} if 0 setdash }
 {pop pop pop Solid {pop []} if 0 setdash} ifelse } def
/BL { stroke gnulinewidth 2 mul setlinewidth } def
/AL { stroke gnulinewidth 2 div setlinewidth } def
/PL { stroke gnulinewidth setlinewidth } def
/LTb { BL [] 0 0 0 DL } def
/LTa { AL [1 dl 2 dl] 0 setdash 0 0 0 setrgbcolor } def
/LT0 { PL [] 0 1 0 DL } def
/LT1 { PL [4 dl 2 dl] 0 0 1 DL } def
/LT2 { PL [2 dl 3 dl] 1 0 0 DL } def
/LT3 { PL [1 dl 1.5 dl] 1 0 1 DL } def
/LT4 { PL [5 dl 2 dl 1 dl 2 dl] 0 1 1 DL } def
/LT5 { PL [4 dl 3 dl 1 dl 3 dl] 1 1 0 DL } def
/LT6 { PL [2 dl 2 dl 2 dl 4 dl] 0 0 0 DL } def
/LT7 { PL [2 dl 2 dl 2 dl 2 dl 2 dl 4 dl] 1 0.3 0 DL } def
/LT8 { PL [2 dl 2 dl 2 dl 2 dl 2 dl 2 dl 2 dl 4 dl] 0.5 0.5 0.5 DL } def
/P { stroke [] 0 setdash
  currentlinewidth 2 div sub M
  0 currentlinewidth V stroke } def
/D { stroke [] 0 setdash 2 copy vpt add M
  hpt neg vpt neg V hpt vpt neg V
  hpt vpt V hpt neg vpt V closepath stroke
  P } def
/A { stroke [] 0 setdash vpt sub M 0 vpt2 V
  currentpoint stroke M
  hpt neg vpt neg R hpt2 0 V stroke
  } def
/B { stroke [] 0 setdash 2 copy exch hpt sub exch vpt add M
  0 vpt2 neg V hpt2 0 V 0 vpt2 V
  hpt2 neg 0 V closepath stroke
  P } def
/C { stroke [] 0 setdash exch hpt sub exch vpt add M
  hpt2 vpt2 neg V currentpoint stroke M
  hpt2 neg 0 R hpt2 vpt2 V stroke } def
/T { stroke [] 0 setdash 2 copy vpt 1.12 mul add M
  hpt neg vpt -1.62 mul V
  hpt 2 mul 0 V
  hpt neg vpt 1.62 mul V closepath stroke
  P  } def
/S { 2 copy A C} def
end
}
\begin{picture}(3600,2160)(0,0)
\special{"
gnudict begin
gsave
50 50 translate
0.100 0.100 scale
0 setgray
/Helvetica findfont 100 scalefont setfont
newpath
-500.000000 -500.000000 translate
LTa
600 251 M
0 1858 V
LTb
600 251 M
63 0 V
2754 0 R
-63 0 V
600 363 M
31 0 V
2786 0 R
-31 0 V
600 511 M
31 0 V
2786 0 R
-31 0 V
600 587 M
31 0 V
2786 0 R
-31 0 V
600 623 M
63 0 V
2754 0 R
-63 0 V
600 734 M
31 0 V
2786 0 R
-31 0 V
600 882 M
31 0 V
2786 0 R
-31 0 V
600 958 M
31 0 V
2786 0 R
-31 0 V
600 994 M
63 0 V
2754 0 R
-63 0 V
600 1106 M
31 0 V
2786 0 R
-31 0 V
600 1254 M
31 0 V
2786 0 R
-31 0 V
600 1330 M
31 0 V
2786 0 R
-31 0 V
600 1366 M
63 0 V
2754 0 R
-63 0 V
600 1478 M
31 0 V
2786 0 R
-31 0 V
600 1626 M
31 0 V
2786 0 R
-31 0 V
600 1701 M
31 0 V
2786 0 R
-31 0 V
600 1737 M
63 0 V
2754 0 R
-63 0 V
600 1849 M
31 0 V
2786 0 R
-31 0 V
600 1997 M
31 0 V
2786 0 R
-31 0 V
600 2073 M
31 0 V
2786 0 R
-31 0 V
600 2109 M
63 0 V
2754 0 R
-63 0 V
600 251 M
0 63 V
0 1795 R
0 -63 V
1163 251 M
0 63 V
0 1795 R
0 -63 V
1727 251 M
0 63 V
0 1795 R
0 -63 V
2290 251 M
0 63 V
0 1795 R
0 -63 V
2854 251 M
0 63 V
0 1795 R
0 -63 V
3417 251 M
0 63 V
0 1795 R
0 -63 V
600 251 M
2817 0 V
0 1858 V
-2817 0 V
600 251 L
LT0
656 1937 M
57 -9 V
56 -15 V
56 -20 V
57 -25 V
56 -28 V
56 -31 V
57 -34 V
56 -36 V
56 -39 V
57 -40 V
56 -43 V
56 -45 V
57 -47 V
56 -50 V
56 -53 V
57 -57 V
56 -63 V
56 -73 V
57 -85 V
56 -111 V
56 -189 V
57 -99 R
56 168 V
57 51 V
56 19 V
56 5 V
57 -3 V
56 -9 V
56 -14 V
57 -19 V
56 -21 V
56 -23 V
57 -25 V
56 -28 V
56 -29 V
57 -31 V
56 -33 V
56 -34 V
57 -36 V
56 -38 V
56 -39 V
57 -42 V
56 -44 V
56 -46 V
57 -48 V
56 -49 V
56 -50 V
57 -59 V
43 -17 V
LT2
656 1901 M
57 -6 V
56 -9 V
56 -13 V
57 -16 V
56 -21 V
56 -23 V
57 -28 V
56 -31 V
56 -35 V
57 -38 V
56 -43 V
56 -45 V
57 -50 V
56 -53 V
56 -57 V
57 -61 V
56 -64 V
56 -68 V
57 -71 V
56 -75 V
56 -79 V
57 -83 V
56 -86 V
57 -90 V
56 -94 V
56 -97 V
57 -99 V
56 -103 V
56 -112 V
stroke
grestore
end
showpage
}
\put(2008,51){\makebox(0,0){$ q $ [ fm$^{-1}$] }}
\put(100,1180){%
\special{ps: gsave currentpoint currentpoint translate
270 rotate neg exch neg exch translate}%
\makebox(0,0)[b]{\shortstack{ $u_0(q)$ [fm $^{3/2}$] }}%
\special{ps: currentpoint grestore moveto}%
}
\put(3417,151){\makebox(0,0){5}}
\put(2854,151){\makebox(0,0){4}}
\put(2290,151){\makebox(0,0){3}}
\put(1727,151){\makebox(0,0){2}}
\put(1163,151){\makebox(0,0){1}}
\put(600,151){\makebox(0,0){0}}
\put(540,2109){\makebox(0,0)[r]{10}}
\put(540,1737){\makebox(0,0)[r]{1}}
\put(540,1366){\makebox(0,0)[r]{0.1}}
\put(540,994){\makebox(0,0)[r]{0.01}}
\put(540,623){\makebox(0,0)[r]{0.001}}
\put(540,251){\makebox(0,0)[r]{0.0001}}
\end{picture}
\caption{ The ghost state ($n=0$) in momentum space. 
The solid (dashed) line is $u^Y_0(q)$ ($u^G_0(q)$).
The disconnection of the solid line causes from change of sign.
}
\label{FIG3.2}
\end{center}
\end{figure}

The ghost state is shown in Fig.\ref{FIG3}. For our calculation the 
eigen value $\gamma _0 $ of Eq. (\ref{eigen}) is not exactly equal to one, but 
0.937 (if it is  renormalized by the abovementioned 94.9\%, $\gamma_0$= 0.987).  
The solid line is the ghost state $u_0^Y$ 
calculated from our Yakubovsky solution 
$\phi_\alpha$, comparing  to the 
dashed line from usual Gaussian function, 
\begin{eqnarray}
u^G_0 (r) = 
\left( { {128~~\omega _{\alpha N} ^3 } \over \pi} \right) ^{1/4} 
\exp (- \omega_{\alpha N} r ^2 )
\end{eqnarray}
with $\omega_{\alpha N} = \Omega \times ( 4 \cdot 1 )/ (4+1)  $ 
where $\Omega $ is a common shell model mode (0.275 fm $^{-2}$)\cite{Kircher}.
They are  normalized by $\int_0 ^\infty u^2_n(r) r^2 dr =1$.
In the short range our ghost state $u^Y_0$ is smaller than $u^G_0$.   
The repulsive core of realistic potentials reflects in this range.  
This  behavior is similar to that of  correlation functions\cite{Nogga}.
Beyond 4 fm our $u^Y_0$ is bigger than $u^G_0$ because in general the Gaussian 
function is more quickly decreasing than exponential one. 
We also  show  them in the momentum space (see Fig.\ref{FIG3.2}). 
Here the repulsive core manifests itself by a node at $\approx$ 2 fm $^{-1}$ which is absent in  
$u_0^G$.
Overall they agree well.  

\begin{figure}[hbtp]
\begin{center}
\input{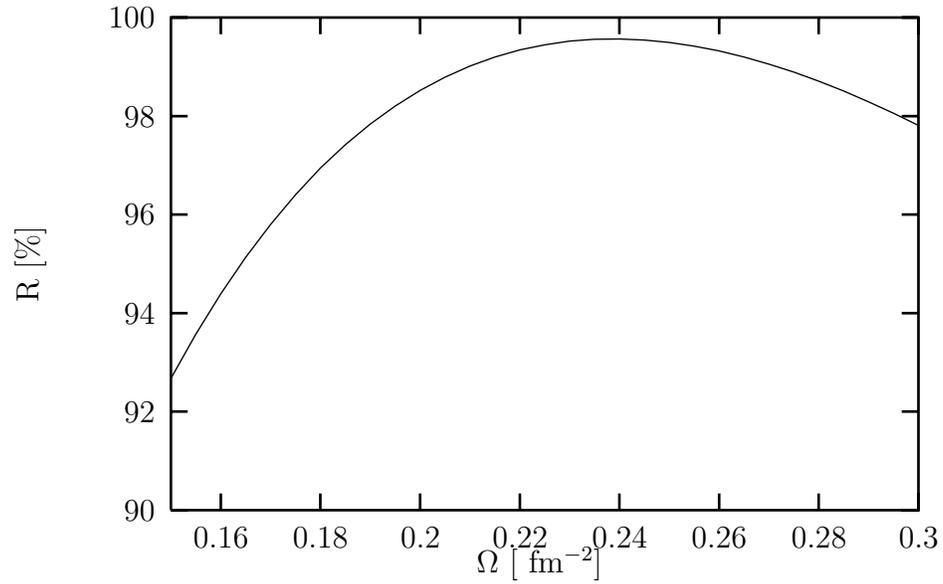}
\caption{ The percentage $R$ of the ghost state as a  
function of oscillator parameter $\Omega$. 
}
\label{FIG4}
\end{center}
\end{figure}

To find the most realistic width parameter $\Omega$ we optimize $R = \vert 
< u^Y_0 \vert u ^G _0 \{\Omega \} > \vert ^2  \times 100 $ [\%]  
in Fig.(\ref{FIG4}). We could recommend $\Omega = 0.24 $[fm$^{-2}$] of 
the Gaussian width parameter which is similar to $\Omega$=0.275 [fm$^{-2}$] \cite{Kircher}.

\begin{figure}[hbtp]
\begin{center}
\input{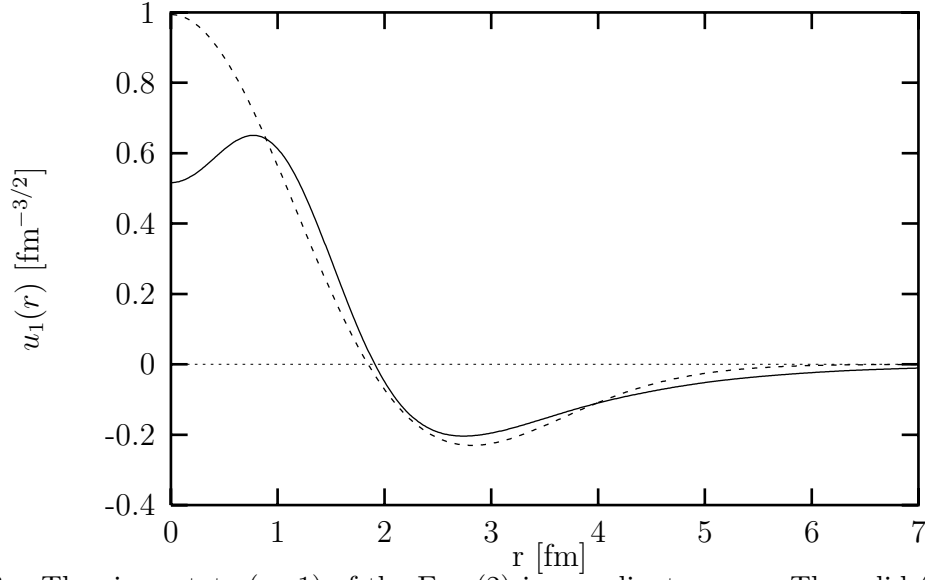}
\caption{ The eigen  state (n=1)  of the Eq. (\ref{eigen})  in coordinate space.
The solid (dashed) line is $u_1^Y(r)$ ($u_1^G(r)$).
}
\label{FIG5}
\end{center}
\end{figure}

In table \ref{table} we summerize the biggest eigenvalues in Eq. (\ref{eigen}). 
Analytically we find in the 
Gaussian case  eigenvalues $\gamma_n = (-4)^{-(2n +0)}$ , $n=0,1,2, \dots$\cite{Zaikin}. 
The realistic NK has got a similar spectrum. 
We compare the states  $u_1^Y$ and $u_1^G$ for n=1 
in Fig. \ref{FIG5}. 
It is remarkable that in this case the realistic ghost state has more structure though the 
eigenvalues are very similar.

The matrix traces $Tr[ {\cal N}_0$] are given 
\begin{eqnarray}
Tr [{\cal N}_0] ^ G &=& \sum _{n=0}^ \infty \gamma _n =\sum _{n=0}^\infty 
\left( { 1 \over 16} \right) ^n = { 16 \over 15 } =1.0666 \dots
\cr
Tr [ {\cal N }_0 ] ^Y &=& \int _0 ^\infty {\cal N}_0( q, q ) q^2 dq =1.0125   
\end{eqnarray}
If the wavefunction of $\alpha$ particle is renormalized by only $j=1/2^+$,
$\int _0 ^ \infty p^2 dp \tilde {\cal N } ( p,p) =1$ ( 0.949 : original norm )
we get $Tr [ {\cal N }_0 ] ^Y$= 1.0666 which must exactly be the  number of the Gaussian form. 
Because it is analytically proved that the trace $Tr [{\cal N}_0]$ does not depend what 
kinds of the $\alpha$ wave function we choose:
\begin{eqnarray}
Tr [ {\cal N}_0 ] &=& \int_0 ^\infty \left[ {1 \over 2 } \int _{-1}^ 1 d \cos \theta 
\tilde {\cal N} ( \sqrt{ { 17 \over 16 } + { { \cos \theta } \over 2} } q ,
                  \sqrt{ { 17 \over 16 } + { { \cos \theta } \over 2} } q )
\right] q^2 dq 
\cr
&=& { 1 \over 2 } \int_{-1}^1 d \cos \theta { 1 \over \sqrt{ { 17 \over 16} + 
{ { \cos \theta } \over 2 } } ^3} = { 16 \over 15}.
\end{eqnarray}

\begin{figure}[hbtp]
\begin{center}
\mbox{\epsfysize=80mm \epsffile{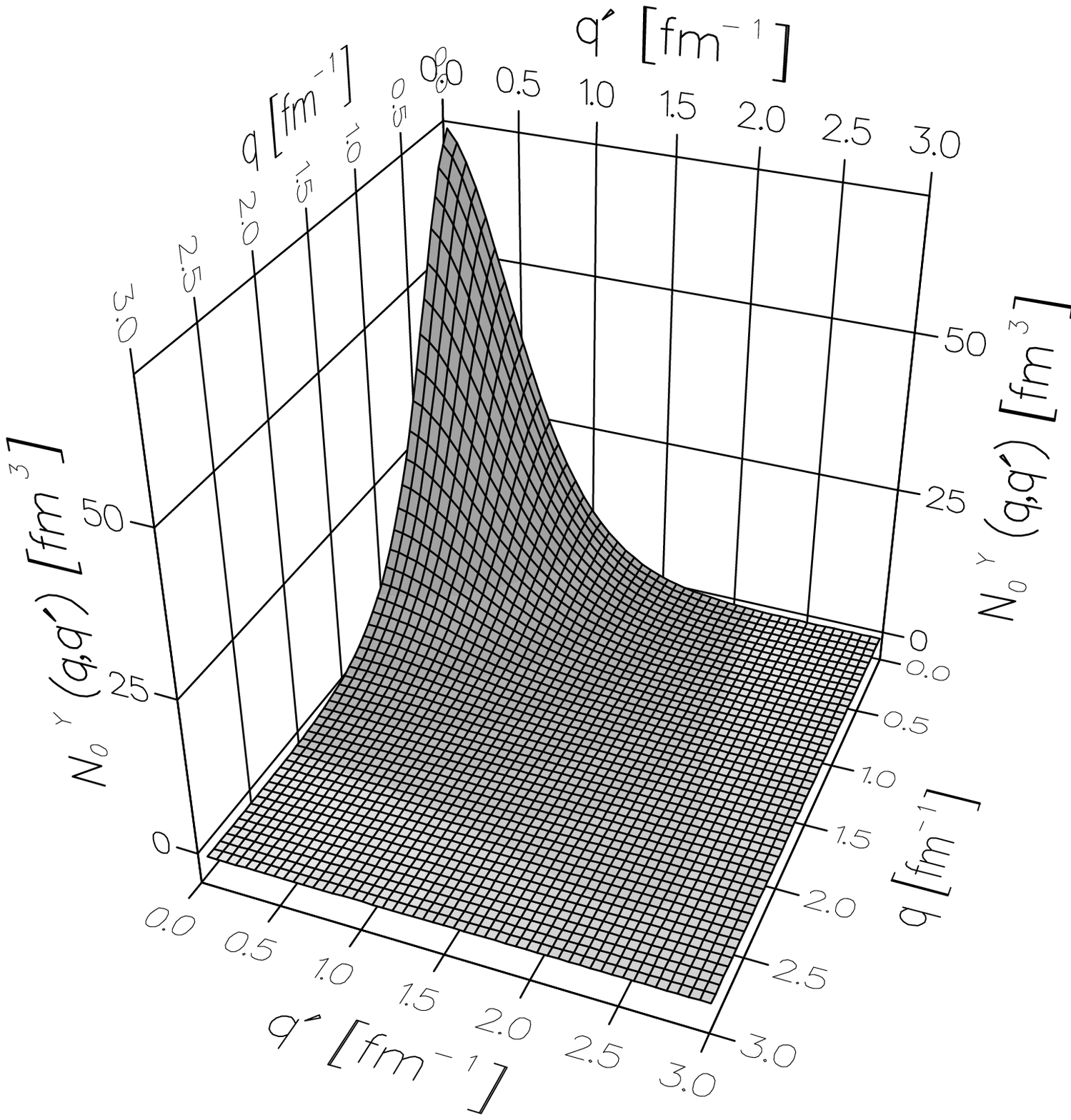}}
\caption{ Norm kernel ${\cal N}_0 ^ Y$.  }
\label{FIG7}
\end{center}
\end{figure}

\begin{figure}[hbtp]
\begin{center}
\mbox{\epsfysize=80mm\epsffile{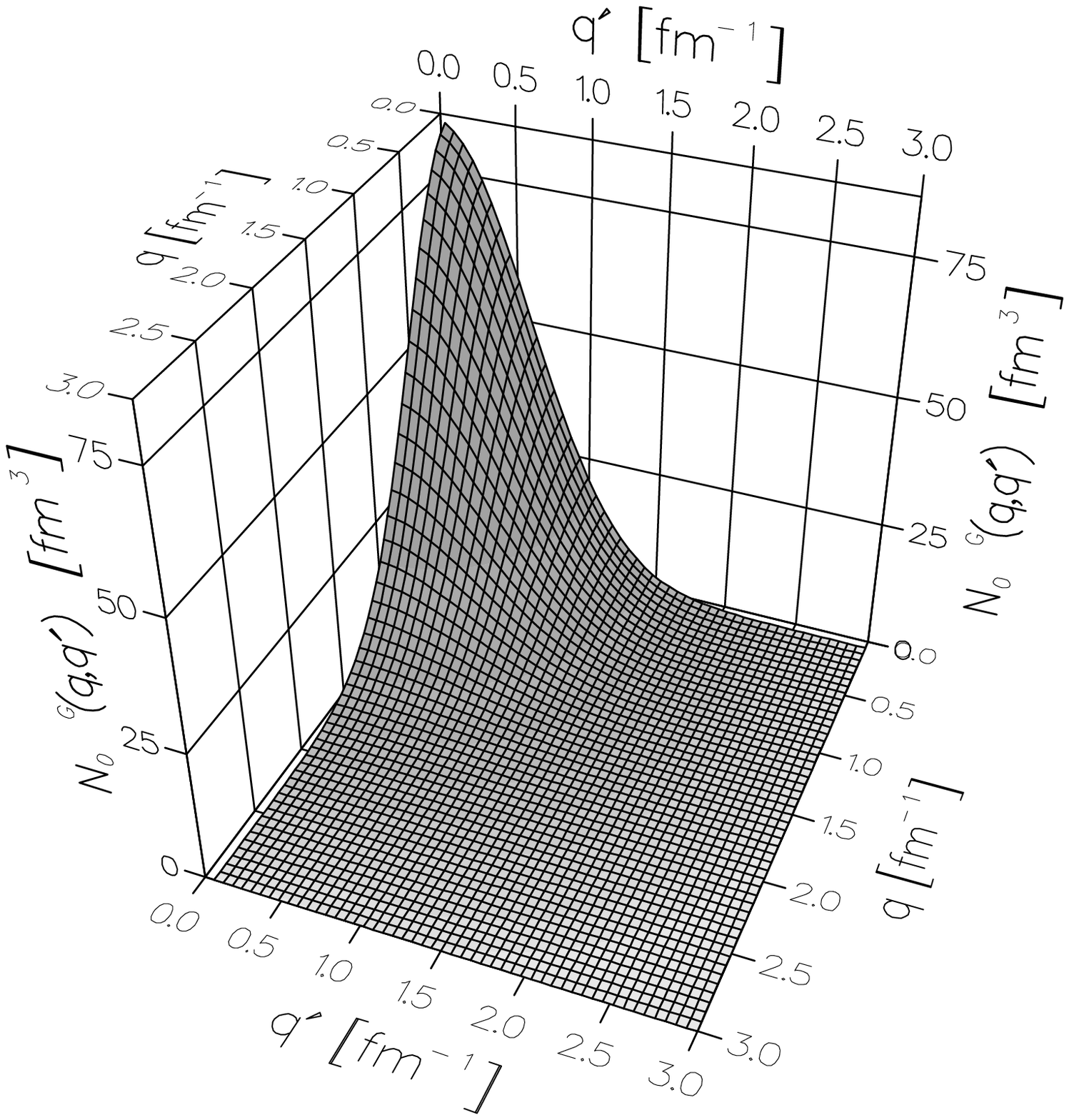}}
\caption{  Norm kernel ${\cal N}_0^G$. }
\label{FIG8}
\end{center}
\end{figure}

\begin{figure}[hbtp]
\begin{center}
\mbox{\epsfysize=80mm\epsffile{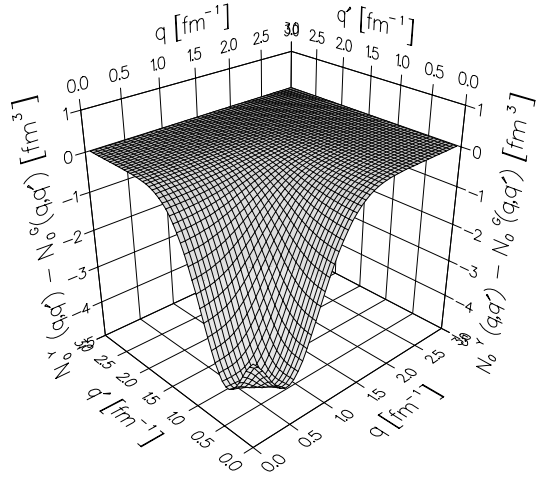}}
\caption{ The difference of the norm kernels(${\cal N}_0^ Y -{\cal N}_0 ^G$). }
\label{FIG9}
\end{center}
\end{figure}

We illustrate both NKs (Fig. \ref{FIG7} 
for ${\cal N}_0^Y$ and  Fig. \ref{FIG8} for ${\cal N}_0 ^Y$). 
The Gaussian case is analytically given, 
\begin{eqnarray}
{\cal N}_0 (q, q') = { 32 \over {qq'}}
\sqrt{ 1  \over{6 \pi \Omega } } \exp \left[ {17 \over {48 \Omega}} (q ^2 + {q'} ^2 ) \right] 
\sinh \left[ { 1 \over { 3 \Omega} } q q' \right]
\end{eqnarray}
The shape is so similar that the difference (${\cal N}_0 ^Y - {\cal N}_0^G$) is also shown in Fig. \ref{FIG9}.

Although there is only a single ghost state in $\alpha$-N system, in general, 
the cluster-cluster effective interaction in light nuclei has
a lot of ghost states.
In this simple case we could find some remarkable differences in the eigen state (n=1) and the 
eigenvalue for n=2 which might effect RGM calculations of  systems with A $>$ 5.
For a most probable case such a Pauli blocking will be 
applied to the   $\alpha$-n-n three-body model 
system. There are already some applications\cite{PRC61.2000,Aoyama}  by using some 
Pauli methods. 

It will be important benchmark calculations for more nucleons system to 
look into the ghost states using rigorous solutions\cite{Schiavilla} 
from  Few-Body Physics. 

Note that here we discuss the Pauli forbidden state which is different from the spurious state of 
the Faddeev calculations\cite{Jensen,SPURIOUS}. The naming of spurious state has been used a lot in many places, even if 
a cluster model has no inner structure the ghost states appear in the model  
and they are interpreted kinds of spurious states.
We should not confuse spurious states caused from Faddeev 
decomposition\cite{Jensen,SPURIOUS,Be9lambda}. 
In this paper we simply take the 
physical Yakubovsky solution of $\alpha$ particle 
to test quantitatively how precise the former Gaussian  norm kernel is.

\bigskip

\acknowledgements
This article is dedicated to the occasion for 60th birthday of Prof. W. Gl\"ockle.  
One of authors (H.K.) 
would like to thank Prof. H. Wita\l a and  Dr. J. Golak for their fruitful 
discussion during my stay in  Cracow. 
This work was supported by the Deutsche Forschungsgemeinschaft (H.K. and A.N.)  and 
The numerical calculations have been performed on the CRAY T90 of the 
John von Neumann Institute for Computing 
in J\"ulich, Germany.

\begin{table}
\caption{\label{table}
Eigen values the norm kernel. 
}
\begin{tabular} {c@{\hspace{5mm}}|r@{\hspace{5mm}}|r@{\hspace{5mm}}|r@{\hspace{5mm}}|r@{\hspace{5mm}}}
n&\multicolumn{1}{c}{ $\gamma_n$ } & { ($ \gamma ^ {-1}_n $) of $u^Y_n$ } 
&\multicolumn{1}{c}{  $\gamma_n$ } & { ($\gamma ^ {-1}_n $) of $u^G_n$ \cite{Zaikin} }\\
\hline
0& 0.937  & (1.068)     & 1.00000 &  (1) \\
1& 0.0663 & (15.09)     & 0.06666 & (16) \\
2& 0.00753  & (132.)     & 0.00391 &(256) \\
\end{tabular}
\end{table}


\end{document}